# Current-density-modulated Antiferromagnetic Domain Switching Revealed by Optical Imaging in Pt/CoO(001) Bilayer

Tong Wu[1], Haoran Chen[1], Tianping Ma[2], Jia Xu[3], Yizheng Wu[1,4,5]*

[1.] Department of Physics and State Key Laboratory of Surface Physics, Fudan University, Shanghai 200433, China

[2.] Anhui Key Laboratory of Magnetic Functional Materials and Devices, School of Materials Science and Engineering, Anhui University, Hefei 230601, China

[3.] Department of Physics, School of Physics and Telecommunication Engineering, Shaanxi University of Technology, Hanzhong 723001, China

[4.] Shanghai Research Center for Quantum Sciences, Shanghai 201315, China

[5.] Shanghai Key Laboratory of Metasurfaces for Light Manipulation, Fudan University, Shanghai 200433, China

**Abstract**

Efficient control of antiferromagnetic (AFM) domain switching in thin films is vital for advancing antiferromagnet-based memory devices. In this study, we directly observed the current-driven switching process of CoO AFM domains in the Pt/CoO(001) bilayer by the magneto-optical birefringence effect. The observed critical current density for AFM domain switching remains nearly constant across varying CoO thicknesses, associated with the consistent switching polarity $\boldsymbol{n} \perp \boldsymbol{j}$, suggesting the dominance of the thermomagnetoelastic effect, where $\boldsymbol{n}$ and $\boldsymbol{j}$ stand for Néel vector and current density vector, respectively. Further confirmation comes from a similar switching process with $\boldsymbol{n} \perp \boldsymbol{j}$ observed in the Pt/Al$_2$O$_3$/CoO sample, excluding the contribution of spin current injection. Remarkably, it was also surprisingly observed that the Néel vector can be further switched parallel to the current direction ($\boldsymbol{n} /\!/ \boldsymbol{j}$) at higher current density. Our findings not only enhance the understanding of current-driven AFM domain switching but also present new avenues for manipulating AFM domains.

# I. Introduction

Antiferromagnetic (AFM) spintronics has attracted great attention in the past decade due to the unique characteristics of antiferromagnets over ferromagnets, especially the absence of net magnetization producing no stray field, making them inert to the external fields and capable of serving as stable multilevel information storage media with high density [1-10]. Besides, their intrinsic precession frequency determined by exchange interaction can reach as high as terahertz (THz), allowing for ultra-fast spin dynamics such as THz pulse-induced AFM spin resonance [11,12]. It is noted that the information storage in AFM spintronic devices is encoded in the Néel vector $\boldsymbol{n}$ within AFM domains, making it imperative to image the AFM domains for comprehending their switching behaviors.

Up to now, X-ray magnetic linear dichroism (XMLD) effect-based photoemission electron microscopy (PEEM) has been the prevailing technique for visualizing AFM domains in various AFM materials, which also allows for observing their changes subsequent to the application of external current or magnetic fields [5-6,13-17]. However, PEEM relies on the collected photoelectrons, making it challenging to concurrently capture the temporal evolution of AFM domains in the presence of an external field. Additionally, several other scanning imaging techniques, such as the nitrogen vacancy centre-based diamond microscope and the spin-polarized scanning tunneling microscope have been developed to measure the spin structure of antiferromagnets [18-19]. Yet, these approaches often necessitate a super-flat surface and operate at low imaging speeds, limiting the real-time analysis of AFM domain switching dynamics.

Recently, a novel method has been proposed to image AFM domains based on the magneto-optical birefringence (MOB) effect, utilizing a tabletop commercial Kerr microscope [20-21]. This innovative technique enables wide-field imaging of AFM domains in thin films and has been first validated in NiO(001) and CoO(001) single crystal films grown on MgO(001) substrates [20-21]. More importantly, as a pure optical technique, it is compatible with current pulses and magnetic fields during the measurement process, allowing for the real-time investigation of AFM domain switching behaviors under external fields [22-23].

The implementation of electrical manipulation of AFM domains is crucial in AFM spintronic devices, rendering the realization of current-driven AFM domain switching and understanding the underlying mechanisms profoundly significant. For AFM metals, such as $Mn_2Au$ and CuMnAs, which exhibit a broken local inversion symmetry in their crystal structures, have shown evidence of current-induced AFM domain switching through anisotropic magnetoresistance (AMR) and XMLD-PEEM measurements [3-5,24-26], and the switching mechanism was attributed to the spin-orbit torque (SOT) effect [3-4]. The cases for AFM insulator/heavy metal systems, such as NiO/Pt, $Fe_2O_3$/Pt and CoO/Pt, are much more complicated. It has been proved that spin Hall magnetoresistance (SMR) can be employed to detect the switching of AFM domains in these systems [7,27-28]. However, the measured magnetoresistance signal was then proved not necessarily to have a one-to-one correspondence with the state of AFM domains [29]. Further investigations with domain imaging based on the MOB effect confirmed current-driven switching of AFM domains in NiO/Pt bilayer, verifying the capacity of the MOB effect to study the dynamics of AFM domain switching in the presence of external fields [23,30]. In such systems, the underlying mechanism of current-driven AFM domain switching was initially explained by the antidamping-like SOT effect in NiO/Pt [7,8]. Subsequently, the thermomagnetoelastic effect was proposed to explain the current-driven AFM domain switching in $Fe_2O_3$/Pt [27], and this explanation was further applied to understand the phenomena in NiO/Pt using optical imaging [30].

CoO is a common collinear G-type antiferromagnet with AFM spins aligning along the <117> directions in its bulk phase with Néel temperature ($T_N$) of 291 K [31]. Because the lattice constant of CoO is slightly larger than that of MgO ($a_{CoO}$ = 4.26Å > $a_{MgO}$ = 4.21Å), the epitaxial CoO(001) thin film on the MgO(001) substrate undergoes in-plane compressive strain, leading to an in-plane biaxial spin alignment along the [110] and [$\bar{1}$10] directions and a higher $T_N$ of about 310 K due to strong magneto-elastic coupling [32-33]. The presence of two available in-plane states for AFM domains, combined with the relatively low $T_N$, makes it suitable for the development of energy-efficient information storage devices under ambient conditions. Previous studies have shown AFM domain switching in CoO/Pt, induced either by current or field investigated by SMR effect [28,34-36] or XMLD signal [35,36]. However, current-driven domain switching in CoO has not been confirmed by direct imaging yet, and the exact switching mechanism still remains elusive [28]. In this paper, we demonstrated the current-driven AFM domain switching in Pt/CoO(001) bilayer by optical imaging. Systematic measurements with variable temperatures and CoO thicknesses were also performed. The measured critical current density was found to be almost independent of CoO thickness and the same switching polarity with $\boldsymbol{n} \perp \boldsymbol{j}$ was observed at various temperatures and CoO thicknesses. Combined with the similar reversible switching in the sample with a 4 nm $Al_2O_3$ spacer layer between CoO and Pt layers, it was confirmed that the current-driven switching of CoO AFM domains should be attributed to the thermomagnetoelastic effect. Furthermore, it was surprisingly discovered that AFM domains in CoO can further align parallel to the direction of the current ($\boldsymbol{n} \mathbin{/\mkern-6mu/} \boldsymbol{j}$) at higher current density, which is distinct from their behaviors at lower current density ($\boldsymbol{n} \perp \boldsymbol{j}$). This new finding allows the domain orientation manipulation by the different current densities instead of the current directions.

## II. Experiments

Single-crystalline CoO(001) films were prepared by molecular beam epitaxy in an ultra-high vacuum system with a base pressure of $2\times10^{-10}$ Torr. The CoO layer was grown on the pre-annealed single-crystal MgO(001) substrate by reactive deposition of Co at an oxygen pressure of $4\times10^{-6}$ Torr at room temperature (RT). In order to verify whether the spin current transmitted through the Pt/CoO interface plays a role in the switching process, it is necessary to study the CoO thickness dependence of current-driven switching process. Thus, the CoO layer was grown into a wedge, ranging from 0 to 8 nm within 6 mm in length by moving the sample holder behind a shield during deposition, as shown in Fig. 1(a). The epitaxial growth of the CoO layer is evident from the sharp reflection high energy electron diffraction patterns in Figs. 1(b)-(c), which confirm the lattice relationship of CoO[100](001) // MgO[100](001). Then, the sample was transferred into the magnetron sputtering system and a 3-nm-thick Pt layer was deposited. Finally, through the standard lithography and $Ar^+$ etching processes, the sample was patterned into the 10-um-wide cross bars with the arms of the devices along two easy axes of CoO, i.e. [110] and [$\bar{1}$10].

We then investigated the current-driven AFM domain switching using a commercial Kerr microscope equipped with an optical cryostat, allowing the operation within a temperature range between 77 K and RT, as illustrated in Fig. 1(d). The CoO AFM domains were imaged with blue light incident normal to the sample. The polarization of incident light is parallel to the <100> direction, ensuring the strongest MOB contrast for the AFM domains [20]. To perform the studies on the current-driven AFM domain switching, the current pulses with the duration of 100 were applied. The pulses were applied through a Keithley 6221 current source, and were switched between [110] and [$\bar{1}$10] directions of CoO. The samples were initialized by applying strong currents along a certain direction to form a single AFM domain. Subsequently, the current pulses were applied in the direction perpendicular to the initial currents, gradually increasing the current density. The CoO AFM domains were imaged 1 second after each current pulse, with an exposure time of ~1 second for imaging. After the application of current pulses, the ratios of switched AFM domains were determined from the MOB images at the crossing area within the devices, as indicated by the red dashed box in Fig. 2(a).

## III. Results and discussion

Figures 2(a)-(d) illustrate the AFM domain switching process in an 8-nm-thick CoO device measured at 80 K with current pulses along the -y direction, indicated by the blue arrows. Fig. 2(a) shows the initial state with the Néel vector $\boldsymbol{n}$ of CoO at the center of the device along the y-axis after applying a strong current along the +x

direction. The relationship between the direction of $\boldsymbol{n}$ and the optical contrast was first characterized in the Fe/CoO(001) film, demonstrating orthogonal coupling between the Fe spins and CoO AFM spins [21]. Figs. 2(b)-(d) represent the intermediate states during the switching process after the pulses with increasing current strengths. The states during the switching process with orthogonal current pulses are also presented in Figs. 2(e)-(h). Fig. 2(e) shows the state with $\boldsymbol{n}$ at the cross aligned along the x-axis. Figs. 2(f)-(h) show the intermediate states after the current pulses applied in the bar along the +x direction, indicated by the red arrows. The clear and reversible switching of AFM domain was observed with the polarity of $\boldsymbol{n} \perp \boldsymbol{j}$ after switching. During the switching process, both the domain nucleation and domain wall motion could be observed. Similar measurements were conducted at different temperatures ranging from 80 K to RT. We then determined the relationship between the switched ratio of AFM domain and current density along the y-axis at different temperatures. The switching curves in Fig. 2(i) confirm 100% switching at all temperatures, but shift towards lower current density while maintaining the similar shape as the temperature increases.

We also imaged the AFM domain switching with different CoO thicknesses ($d_{CoO}$), as depicted in Fig. 3 at 80 K. Figs. 3(a)-(d) exhibit the final states of CoO films with different $d_{CoO}$ after current pulses applied along the +x direction, while Figs. 3(e)-(h) show the final states after current pulses applied along the -y direction. The optical contrast of AFM domains increases with increasing $d_{CoO}$, owing to the thickness-dependence of the MOB effect in CoO film [21]. All the CoO films with different thicknesses show the same switching polarity of $\boldsymbol{n} \perp \boldsymbol{j}$, implying the same switching mechanism independent of film thickness. Fig. 3(i) shows the measured switching curves with different $d_{CoO}$ at 80 K, which have very weak thickness-dependence on AFM domain switching. The critical current density of switching curves varies by approximately 1.8% across CoO thicknesses ranging from 3 nm to 8 nm. Even in the case of the 1.8-nm-thick CoO film, the switching current density is only about 4% lower compared to the devices with thicker films.

To quantitatively investigate the impact of temperature and CoO thickness on the current-driven AFM domain switching process, the critical current densities ($j_c$) were extracted from the switching curves with different $d_{CoO}$ at various temperatures as illustrated in Figs. 4(a)-(b), where $j_c$ signifies the current density with a 50% switched ratio. For the device with 1.8-nm-thick CoO , the contrast of AFM domains is too weak to be distinguished above 125 K. For the devices with $d_{CoO}$ ranging from 3 nm to 5.4 nm, the switched ratio at high temperatures cannot exceed 50%, possibly due to the current-induced random thermal excitation overcoming anisotropy barrier, resulting in a multi-domain state. At a fixed temperature, $j_c$ remains nearly independent of $d_{CoO}$. This observation suggests an origin of AFM domain switching that does not rely on the interface, which would typically weaken with increased thickness. Conversely, with a fixed $d_{CoO}$, $j_c$ exhibits a linear decrease as the temperature increases as depicted in Fig. 4(b). It was reported that the Joule-heating-induced temperature increase ΔT is proportional to $j^2$ [27], and the magnetic anisotropy of the AFM CoO film is also reduced due to the increasing temperature. Thus, the linear relationship between $j_c$ and temperature here should not only be attributed to the Joule-heating-induced temperature increase, but may also arise from the reduction of the switching energy barrier of CoO due to thermal excitation. Fig. 4(b) also shows that the relationships between critical current density and temperature for different $d_{CoO}$ are concentrated within a narrow range, due to the weak $d_{CoO}$-dependence on the switching process shown in Fig. 4(a). While decreasing the temperature, $j_c$ increases from ~$3.1\times10^{11}$A/m$^2$ at 295 K to ~$13.3\times10^{11}$A/m$^2$ at 80 K.

Next, we will further discuss the physical mechanisms governing current-driven AFM domain switching in Pt/CoO bilayer. Regarding the current-driven AFM domain switching in the heterostructure of heavy metal/AFM insulator, two mechanisms have been proposed. One is the antidamping-like spin-orbit torque (SOT) resulting from the spin current generated in the heavy metal layer [7-8], capable of aligning the Néel vector parallel to the current direction ($\boldsymbol{n}\;//\;\boldsymbol{j}$) as depicted in Fig. 4(c). The other mechanism involves the Joule-heating-induced magnetoelastic coupling effect induced by the current pulse [27-28,30]. Because of the Joule heating generated by the applied current pulse, the current path experiences a higher temperature compared to the surrounding substrate. Consequently, a greater thermal expansion occurs at the central current path, leading to a net tensile strain perpendicular to the current direction, as illustrated in Fig. 4(d). Such tensile strain can induce in-plane uniaxial anisotropy in the AFM layer due to the magnetoelastic coupling effect, which can further drive the AFM spins towards the easy axis during the cooling process after the current pulse. The magnetic energy introduced by the Joule-heating-induced magnetoelastic coupling scales as $E \propto -\lambda(\bm{n}\cdot\bm{j})^2$ [28], where $\lambda$ stands for the magnetoelastic coefficient. Therefore, a positive $\lambda$ signifies the parallel alignment of the final AFM spins with the current direction, i.e. $\bm{n}//\bm{j}$, whereas a negative $\lambda$ results in AFM spins perpendicular to the current, indicated by the relationship $\bm{n}\perp\bm{j}$. The current-induced switching effect in Pt/CoO cannot be explained by the antidamping-like SOT effect, since the observed final state of AFM domains follows the relationship of $\bm{n}\perp\bm{j}$, which is in line with the thermomagnetoelastic switching mechanism for a negative magnetoelastic constant of CoO [28]. Our findings, exploring current-induced switching as a function of temperature and $d_{CoO}$, strongly support the thermomagnetoelastic switching mechanism. Notably, the critical current density remains independent of $d_{CoO}$, excluding the mechanism associated with the SOT effect, which typically exhibits significant thickness dependence. However, the thermal-induced strain is mainly determined by the insulating substrate and the metallic current line, and has no direct correlation with the CoO film. In this case, the thermomagnetoelastic switching mechanism should have no dependence on the CoO thickness.

In order to further verify the thermomagnetoelastic switching mechanism, we conducted the simulations using COMSOL following the methods described in Ref. [27-28,30]. The modeling was performed based on the crossbar device with the real dimensions such as the length of 110 μm and the width of 10 μm, as indicated in Fig. 4(e). The thicknesses of Pt and CoO layers were set as 3 nm and 8 nm, respectively. The MgO substrate was modeled with dimensions of 1 mm × 1 mm × 10 μm. We chose the same boundary conditions and material parameters such as the heat capacity, thermal conductivity and heat transfer coefficient as in Ref. [28,30]. A pulse current with a width of 100 μs and an amplitude of 15 mA was injected into the arm along the +x direction at 295 K, then the thermal and strain distributions at the bottom surface of the CoO layer were calculated as a function of time. Fig. 4(e) shows that the temperature in the current path at the time of 100 μs significantly exceeds that of the surrounding substrate, reaching an instantaneous maximum of 320 K, closely approaching the $T_N$ of CoO film. The non-uniform temperature distribution within the film will further generate strain. The lower temperature in the cross area can be attributed to the higher heat transfer rate in the Pt bars perpendicular to the current. However, the relatively lower temperature near the end of the current-carrying electrode is due to the additional heat dissipation boundary at the end of the device, as heat generation mainly occurs within the device. In the actual experimental setup, the end was connected to a larger and thicker electrode area for wire-bonding, resulting in decreased current density and naturally lower temperature in this area. Fig. 4(f) presents the calculated distribution of the net tensile strain defined as $\varepsilon_{xx}$-$\varepsilon_{yy}$, with $\varepsilon_{xx}$ and $\varepsilon_{yy}$ as the strain along the x and y axes, respectively. The positive or negative values of $\varepsilon_{xx}$-$\varepsilon_{yy}$ corresponds to the net tensile strain along the x or y axes, respectively. Thus, the calculated negative net tensile strain shown in Fig. 4(f) can induce a uniaxial anisotropy with the easy axis along the y axis due to the negative magnetoelastic constant. Note that the maximum temperature after the current pulse can reach the $T_N$ of CoO film, thus during the cooling process, the magnetoelastic effect-induced uniaxial anisotropy can drive the AFM spins along the y-axis. At lower experimental temperatures, achieving a sufficiently high sample temperature change for AFM domain switching requires a higher current strength, as shown in Fig. 2(i). Therefore, our numerical simulations further corroborate the dominance of the thermomagnetoelastic effect in CoO/Pt bilayer for AFM domain switching. It should be noted that, in the time-dependent simulations, the strain relaxation rate is much faster than the temperature relaxation rate. For example, while performing the simulation with a current of 35 mA at 295 K, the sample temperature can still remain above 315 K at 0.3 ms after the current pulse, whereas the strain is negligible, being three orders of magnitude smaller than its peak value. Thus, the PM-AFM phase transition occurs under conditions of nearly zero strain distribution, potentially leading to the random distribution of AFM domains due to the four-fold symmetry in the CoO(001) film.

While further increasing the current density, we observed an additional domain switching with the final state of $\boldsymbol{n} /\!/ \boldsymbol{j}$. Fig. 5(a) shows the switching curves and the domain states at several characteristic current densities at the cross area of the device with 8-nm-thick CoO film, as a function of the current density for the current pulses applied along two bars. The white and black contrasts denote the $\boldsymbol{n}$ along the x and y axes, respectively, indicated by the double arrows. For $\boldsymbol{j} /\!/ [\bar{1}10]$(-y axis), the domain switched to the white-contrast state of $\boldsymbol{n} \perp \boldsymbol{j}$ at the critical current density of $j_c \sim 4.6 \times 10^{11}$ A/m$^2$. However, it also reveals an additional domain switching with the final black-contrast state of $\boldsymbol{n} /\!/ \boldsymbol{j}$ at the higher critical current density of $j_{hc} \sim 9.8 \times 10^{11}$ A/m$^2$. For $\boldsymbol{j} /\!/ [110]$ (+x axis), the similar two-step switching behavior can also be observed with the critical current densities of $j_c \sim 4.1 \times 10^{11}$ A/m$^2$ and $j_{hc} \sim 11.0 \times 10^{11}$ A/m$^2$. The current-induced switching for $\boldsymbol{j} /\!/ [110]$ exhibits a lower $j_c$ but a higher $j_{hc}$. It is possible that the CoO film has a weak in-plane uniaxial anisotropy with easy axis along $[\bar{1}10]$, which can be induced by the possible small miscut of MgO(001) substrate [37].

The results in Fig. 5(a) provide an unconventional opportunity to switch the AFM domains by different current densities. We applied two distinct currents along the x-axis, with the densities $j$ of $6 \times 10^{11}$ A/m$^2$ ($j_c < j < j_{hc}$) and $12 \times 10^{11}$ A/m$^2$ ($j > j_{hc}$), and measured the MOB signal at the cross area. Fig. 5(b) clearly shows the alternating changes in domain contrast induced by these two varied currents. Figs. 5(c)-(f) show the representative domain images subsequent to the current pulses with alternating strengths, which clearly demonstrate the alternative switching by the current pulses with different densities. This result highlights the potential for achieving reversible switching of AFM domains by adjusting the current density in a simple two-terminal device, deviating from the traditional method of altering the current direction in a four-terminal device with a crossbar structure [28].

It still remains unclear about the mechanism for the CoO AFM domain switching into the state of $\boldsymbol{n}\,/\!/\,\boldsymbol{j}$ at high current density. Our numerical simulations indicate that the current pulses consistently induce net tensile strain perpendicular to the current direction, regardless of current density. The temperature tends to stabilize quickly, with an increase of less than 3 K at 2 ms after the current pulse compared to the initial temperature, even with a strong current of 35 mA. The transient strain stabilizes even faster, almost disappearing within 1 ms. Thus, the ordinary magnetoelastic effect can only account for the AFM domain switching into the state of $\boldsymbol{n}\perp\boldsymbol{j}$. It is noted that the antidamping-like SOT effect can induce the final state with $\boldsymbol{n}\,/\!/\,\boldsymbol{j}$ [7-8], suggesting the possibility that the additional switching at higher current density may be influenced by the spin current from the Pt layer. To investigate the influence of the spin current from the Pt layer, we inserted a 4-nm-thick $Al_2O_3$ layer into the Pt/CoO bilayer, aimed at blocking the injection of spin current from the Pt layer to the CoO layer. As a result, this configuration should effectively eliminate the possible SOT effect in the system, as illustrated in Fig. 6(a). We conducted similar current-driven AFM domain switching measurements on a Pt(3 nm)/$Al_2O_3$(4 nm)/CoO(8 nm) sample. Our measurements first demonstrated that the same current-induced AFM domain switching with $\boldsymbol{n}\perp\boldsymbol{j}$ could be achieved at low current density, as shown in Figs. 6(b)-(c). This result is consistent with the switching mechanism of the thermomagnetoelastic effect. However, while further increasing the current strength, the second current-induced domain switching with $\boldsymbol{n}\,/\!/\,\boldsymbol{j}$ could be also observed. Remarkably, this alternate switching can be triggered by currents of two different densities, as evidenced by the domain images in Figs. 6(d)-(g). Hence, the domain switching at high current density above $j_{hc}$ in the Pt/$Al_2O_3$/CoO sample effectively excludes the likelihood of its origin from the SOT effect. This additional switching might also be attributed to the thermomagnetoelastic effect induced by the current pulse. Our numerical simulations indicate that the current pulse with $j=12\times 10^{11}$ A/m$^2$ can elevate the temperature at the device center to ~470 K from RT, which is far above the $T_N$ of CoO film. Thus the heat conduction process may be nonlinear and anisotropic in the crystalline system, which is beyond the capability of conventional numerical simulation method employed in this study. It should also be noted that when a large current is applied to induce the AFM domain switching with $\boldsymbol{n}/\!/\boldsymbol{j}$, strain could lead to the formation of grid-like defects in the sample oriented along the <100> direction, as shown in Figs. 5(c)-(f) and Figs. 6(e)-(g), potentially restricting the numbers of switching. Although the mechanism of the current-induced switching of CoO AFM domains at high current density remains unclear, our findings demonstrate the possibility to reversibly manipulate the AFM domains only by adjusting the current density instead of changing the direction of the current. This feature can promote the development of a new two-terminal device that has the potential to simplify the complexity of antiferromagnet-based devices.

## IV. Summary

In summary, we systematically investigated the current-driven switching of AFM domains in the Pt/CoO(001) system through direct optical imaging, considering various temperatures, CoO thicknesses, and different strengths of current pulses. We discovered that the critical current density $j_c$ for AFM domain switching is independent of the CoO thickness. Furthermore, different CoO thicknesses and temperatures yield the same switching polarity with $\boldsymbol{n}\perp\boldsymbol{j}$. Our results suggest that the AFM domain switching in CoO/Pt bilayer is dominant by the thermomagnetoelastic effect, and this conclusion is further supported by simulation results. The analogous switching process observed with $\boldsymbol{n}\perp\boldsymbol{j}$ in the Pt/Al$_2$O$_3$/CoO sample further rules out the possibility of the SOT effect contributing to the phenomenon. The observation of additional domain switching with $\boldsymbol{n}/\!/\boldsymbol{j}$ at higher current density, while maintaining the same current direction, indicates the potential to manipulate AFM domains in a two-terminal device solely by adjusting the current density. Our experimental investigation not only offers a deeper understanding of current-driven AFM domain switching but also unveils novel prospects for manipulating AFM domains in AFM spintronic devices.

**Acknowledgements:**

The work was supported by the National Key Research and Development Program of China (Grant No. 2022YFA1403300), the National Natural Science Foundation of China (Grant No. 11974079, No. 12274083 and No. 12221004), the Shanghai Municipal Science and Technology Major Project (Grant No. 2019SHZDZX01), and the Shanghai Municipal Science and Technology Basic Research Project (Grant No. 22JC1400200 and No. 23dz2260100). Jia Xu acknowledges support by National Natural Science Foundation of China (Grant No. 12204295), Natural Science Foundation of Shaanxi Provincial Department of Education (Grant No. 22JK0310), Natural Science Basic Research Program of Shaanxi (Grant No. 2022JQ-017), Shaanxi University of Technology (SLGRCQD2125), Youth Hanjiang Scholar Research Support Fund from Shaanxi University of Technology.

*Note added.*－During the review process, we became aware of the related work of Refs. [38], in which the XMLD-PEEM was utilized to investigate the AFM domain switching in Pt/CoO system induced by the nonuniform current passing through the neighboring arms of a cross bar. It was claimed that both the thermomagnetoelastic effect and SOT effect contribute to the AFM domain switching.

Figures:

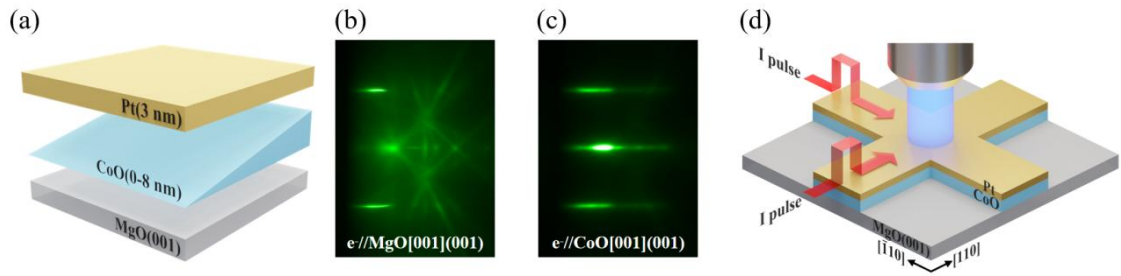

**Fig. 1.** (a) The structure of the Pt/CoO(001) sample with the CoO layer grown into a wedge shape. (b)-(c) RHEED patterns of (b) MgO(001) substrate and (c) 8 nm CoO(001) film. (d) Schematic illustration of the MOB setup with the crossbar structure.

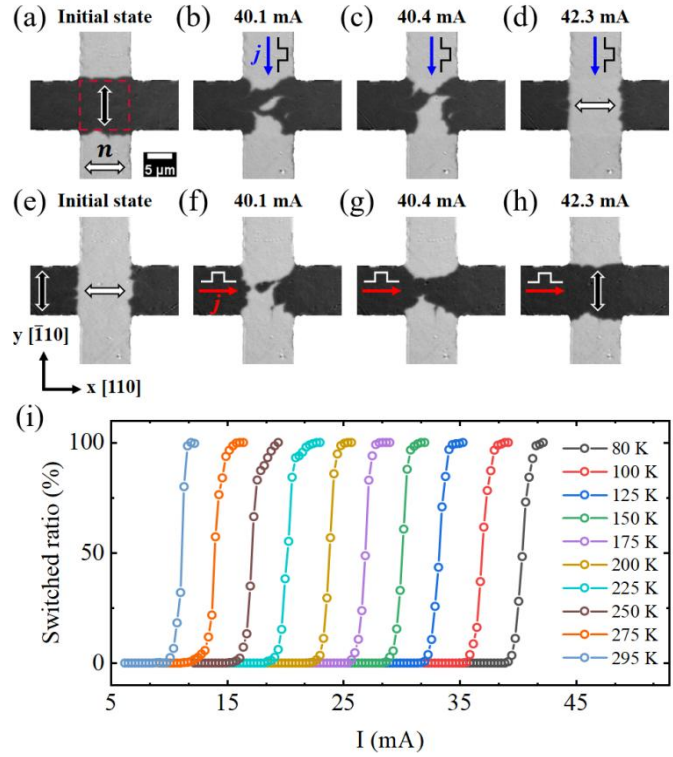

**Fig. 2.** (a) The initial state with $n \parallel$ y-axis at the cross. (b)-(d) Intermediate states after the pulses along the -y direction with increasing current densities. (e) The initial state with $n \parallel$ x-axis in the cross area. (f)-(g) Intermediate states after the pulses along +x direction with increasing current densities. (i) Switching curves of 8 nm CoO at different temperatures.

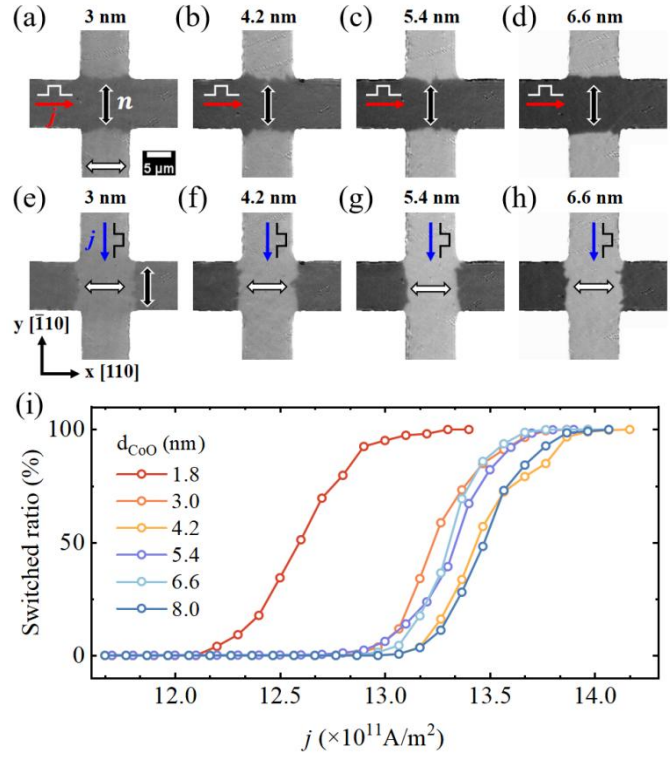

**Fig. 3.** (a)-(d) Final states with $\boldsymbol{n}$ // y-axis, and (e)-(h) final states with $\boldsymbol{n}$ // x-axis at the cross with different CoO thicknesses after the current pulses applied in different bars. The current pulses are indicated by the red or blue arrows in each image. (i) Switching curves at 80 K with different CoO thicknesses.

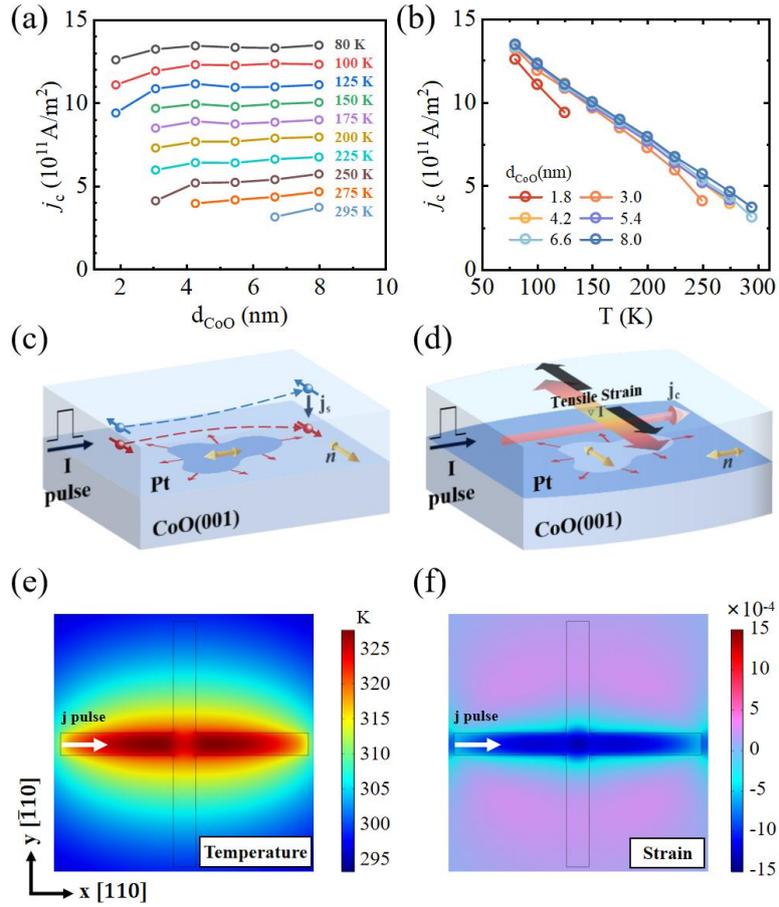

**Fig. 4.** (a) The switching current density $j_c$ as a function of $d_{CoO}$ at different temperatures. (b) The switching current density $j_c$ as a function of temperature with different $d_{CoO}$. (c) Schematic diagram of the SOT-induced switching mechanism. (d) Schematic diagram of the thermomagnetoelastic effect induced switching mechanism. (e) Temperature and (f) strain distribution simulated after applying the current pulse with a width of 100 μs and an amplitude of 15 mA.

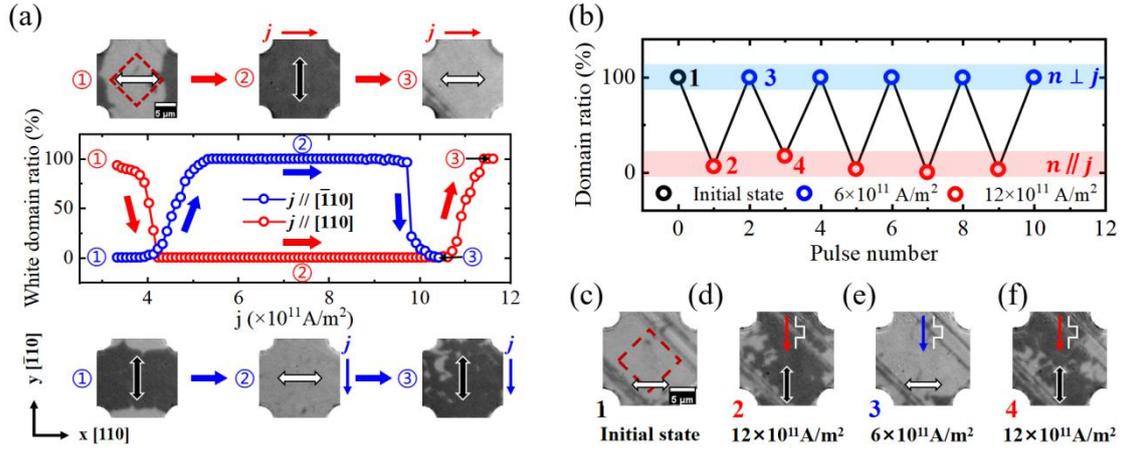

**Fig. 5.** (a) Switching curves for 8 nm CoO measured at 275 K with $j // [\bar{1}10]$ (-y) (blue hollow point line) and $j // [110]$ (x) (red hollow point line), respectively. The insets illustrate the AFM domain states at several characteristic current densities. (b) The ratio of AFM domain states with the AFM spins along [110] direction after applying the current pulses with alternating magnitude. (c)-(f) Initial state and states after current pulses with the current densities of $12 \times 10^{11}$ A/m² (red arrow) and $6 \times 10^{11}$ A/m² (blue arrow), respectively, corresponding with the states marked by the numbers in (b). The evolution of the domain ratio was determined in the center area of the cross marked by the red dashed squares in (a) and (c).

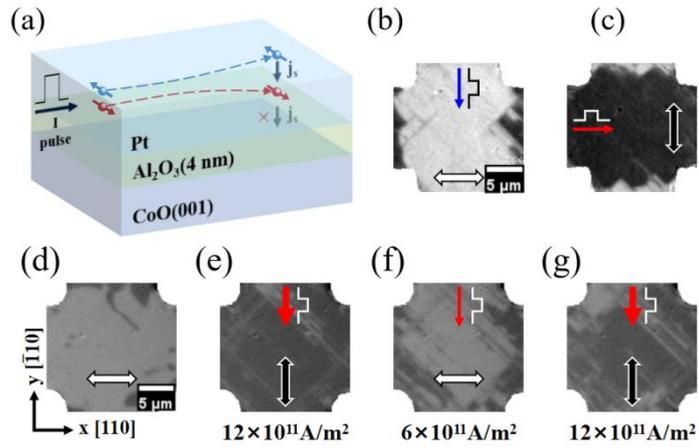

**Fig. 6.** (a) Schematic diagram illustrating the blocking effect of the $Al_2O_3$ interlayer on the injection of spin current from the Pt layer to the CoO layer. (b)-(c) The states after current pulses along (b) -y and (c) +x directions, respectively. (d) Initial state with $\boldsymbol{n}\,/\!/$ x-axis at the device center. (e)-(g) Final states after current pulses with alternating magnitudes.